\documentclass[pra,bibnotes,twocolumn]{revtex4}
\usepackage{graphicx}
\begin{document}
\draft
\def\ds{\displaystyle}
\title{Majorana Edge Modes with Gain and Loss}
\author{C. Yuce}
\address{Department of Physics, Anadolu University, Turkey }
\email{cyuce@anadolu.edu.tr}
\date{\today}
\pacs{ 11.30.Er, 03.65.Vf, 73.21.Cd}
\begin{abstract}
We consider a non-Hermitian generalization of the Kitaev model and study the existence of stable Majorana zero energy modes. We show that they exist in the limit of zero chemical potential even if balanced gain and loss are randomly distributed along the lattice. We show that Majorana zero modes also appear if the chemical potential is different from zero provided that not the full Hamiltonian but the non-Hermitian part of the Hamiltonian is PT symmetric.
\end{abstract}
\maketitle

\section{Introduction}

Non-Hermitian Hamiltonians have attracted a great deal of attention after the paper by Bender and Boettcher, who showed that spectrum of a $\mathcal{PT}$ symmetric non-Hermitian Hamiltonian could be real, where $\mathcal{P}$ and $\mathcal{T}$ operators are parity and time reversal operators, respectively \cite{bender}. More specifically, spectrum of a $\mathcal{PT}$ symmetric non-Hermitian Hamiltonian is real unless non-Hermitian degree exceeds a critical number. If it is beyond the critical number, $\mathcal{PT}$ symmetry is spontaneously broken and the energy spectrum becomes either partially or completely complex. An experiment on a $\mathcal{PT}$ symmetric system with balanced gain and loss was realized in an optical system \cite{deney34}.\\
Topological insulator in $\mathcal{PT}$ symmetric systems is an emergent field in physics \cite{PTop2,PTop3,PTop4,hensch,PTop1,ekl56,cemyuce,cem2}. A topological insulator has gapped energy spectrum in the bulk while it has gapless robust edge states \cite{hasan} (reference therein). Hu and Hughes \cite{PTop2} and Esaki et al.  \cite{PTop3} were the first authors to search for topological phase for non-Hermitian systems. Unfortunately, they found no stable topological states since their system admits complex energy eigenvalues. A Dirac-type non-Hermitian Hamiltonian was considered in \cite{PTop2} and it was concluded that the appearance of the complex eigenvalues is an indication of the absence of the topological insulator phase. A non-Hermitian generalizations of the Luttinger Hamiltonian and Kane-Mele model were considered in \cite{PTop3} and associated topological phase was shown to be unstable. Later, other authors studied a one dimensional non-Hermitian tight binding model \cite{hensch} and non-Hermitian Su-Schrieffer-Heeger (SSH) model with two conjugated imaginary potential located at the edges of the system \cite{PTop1}. They found no stable topological phase in these two systems, too. The first example of topological insulating phase for a non-Hermitian system with real spectrum appeared in the literature recently \cite{cemyuce}. The SSH model with gain and loss impurities located away from edges was shown to admit real spectrum in the topologically nontrivial region. The first experiment in which a topological transition occurs in a lossy non-Hermitian system was performed by Szameit's group \cite{ekl56}. $\mathcal{PT}$ symmetric Floquet topological system was also studied and it was shown that some two dimensional non-Hermitian system may have stable Floquet topological phase \cite{cem2}. \\
 Another interesting system that has recently attracted great attention is the one where Majorana zero modes appear \cite{ettere,majrev1,majrev2,majrev3}. They are interesting since they can be realized in condensed matter systems although its original prediction as elementary particles was made in high energy physics. The particles creation and annihilation operators are equal and consequently Majorana fermions are their own antiparticles. Of special interest is Majorana zero modes, which are Majorana particles with exactly zero energy. A fermion can be thought of as composed of two Majorana fermions and an interesting situation arises when two Majorana zero modes are spatially separated from each other. Such a highly delocalized zero modes in the topologically nontrivial phase are robust against local defects and disorder. More than a decade ago, Kitaev proposed an exactly solvable model with open boundaries \cite{alexi}. It is a one-dimensional tight-binding model for spinless fermions in the presence of p-wave superconducting pairing. He found unpaired Majorana zero modes that commute with the Kitaev Hamiltonian. They are localized near the edges and decay exponentially away from the ends. Note that the Kitaev model can be in two distinct phases depending on the model parameters: the trivial non-topological phase and the nontrivial topological phase. Recently, Kitaev model with a gain at one edge and a loss at the other edge was investigated \cite{ptkitaev1}. The non-Hermitian Kitaev model was made $\mathcal{PT}$ symmetric by assuming that superconducting phase is purely  imaginary. It was shown that the system may admit real spectrum in both topologically trivial and non-trivial regions. The link between exceptional points and Majorana modes has recently been investigated in \cite{agu}.\\
In this paper, we explore Majorana zero modes in a non-Hermitian system. We consider a non-Hermitian generalization of the well known Kitaev model by introducing gain and loss into the standard Kitaev chain. We write our Hamiltonian in Majorana basis and study the existence of stable Majorana zero energy modes. The time-reversal symmetry is broken in the Kitaev model and hence our system is not $\mathcal{PT}$ symmetric. We suppose that the non-Hermitian part of our Hamiltonian is $\mathcal{PT}$ symmetric. In this way, we show that the system admits real energy spectrum and Majorana zero modes are available. We prove that Majorana zero modes located exactly at edges appear when the chemical potential is zero even if balanced gain and loss are randomly distributed along the chain.

\section{Majorana Zero Energy Modes}

We begin with a one dimensional Kitaev model with $N$ sites. The model contains a tight binding chain of spin polarized electrons that can only hop to nearest-neighbor sites. Neighboring electrons can also form Cooper pairs with a superconducting pairing potential term $\ds{\Delta>0}$. In addition to the standard Kitaev model, we introduce gain and loss into the system. A non-Hermitian generalization of the Kitaev Hamiltonian reads
\begin{eqnarray}\label{kiraevham}
H=\sum_{j=1}^{N-1} -T~{a}^\dagger_j a_{j+1}+\Delta a_{j} a_{j+1}+h.c.\nonumber\\
-\mu\sum_{j=1}^{N}  ({a}^\dagger_j  a_{j}-\frac{1}{2})+i\sum_{j=1}^{N} g_{j} {a}^\dagger_{j } a_{j}
\end{eqnarray}
where $\ds{{a}_j}$ and $\ds{{a}^\dagger_j}$ are the electron annihilation and creation operators localized at site $j$, respectively, $
T$ is the constant hopping amplitude, $\ds{\mu}$ is the chemical potential and $\ds{g_{j}}$ are site-dependent non-Hermitian strength. The parameters $\ds{T,\Delta,\mu}$ and $\ds{g_j}$ are all real valued. Note that positive (negative) values of the non-Hermitian strength describe gain (loss). We impose two conditions on the non-Hermitian strength. First, gain and loss are assumed to be balanced in our system. Secondly, we suppose that no gain and loss occur at the two edges of the chain. These are given by
\begin{equation}\label{kfha?}
\sum_{j=1}^N g_j=0~;~~~~g_1=g_N=0
\end{equation}
The reasons why we impose these two conditions will be clear below. In the figure-1, we illustrate our system. The shaded circles represent lattice sites with either gain or loss. In the figure, we consider a Kitaev chain with $N=12$ and $\ds{g_j=g_0(\delta_{j2}-\delta_{j,11})}$, where $g_0$ is a constant. This means that gain is introduced at $j=2$ site and particles are lost at $j=11$ site. The Kitaev model exhibits a topologically nontrivial (top) and trivial (bottom) superconducting phase. In the topological superconductor phase, there are localized Majorana zero modes. They are exactly localized at two edges if the chemical potential is zero. If it is different from zero, the corresponding wave function decays exponentially along the chain. In this paper, we explore whether stable Majorana zero modes are still available if gain and loss are introduced into the system.\\
Let us first discuss the symmetry property of this Hamiltonian. We emphasize that time-reversal symmetry is broken since we only consider one value for the spin projection. The Kitaev Hamiltonian is not also symmetric under combined $\mathcal{PT}$ operation. Although the total Hamiltonian is not $\mathcal{PT}$ symmetric, one can demand that the non-Hermitian part of it is $\mathcal{PT}$ symmetric. In this case, the non-Hermitian strength should be chosen appropriately to satisfy this condition, too. However, as we will show below, stable Majorana zero energy modes appear even if the gain and loss are randomly placed along the lattice.\\
The Hermitian Kitaev Hamiltonian can be rewritten in the form
 \begin{eqnarray}\label{kiraevhammatrix}
H=\frac{1}{2}\sum_{k=0}^{\infty}    (a_k^{\dagger},a_{-k})  \mathcal{H} \left(\begin{array}{cc}a_k  \\a_{-k}^{\dagger} \end{array}     \right)
\end{eqnarray}
where $\ds{ \mathcal{H} }$ is the so-called Bogoliubov-de Gennes Hamiltonian. In the absence of gain and loss, this Hamiltonian has the form
\begin{eqnarray}\label{kiraevhammatrix2}
\mathcal{H}=-(2T\cos {k}+\mu)\sigma_z+2\Delta \sin{k} ~\sigma_y
\end{eqnarray}
where $\sigma_y, \sigma_z$ are Pauli matrices. If the system is infinitely long, the excitation spectrum of the Bogoliubov-de Gennes Hamiltonian is given by $\ds{
E_k= \mp\sqrt{(2T\cos {k}+\mu)^2+4\Delta^2 \sin^2{k} }}$. The two energy eigenvalues are seperated by a gap which closes when $\ds{\mu=\mp 2T}$. This is the signal of topological phase transition. If the chain is finite, one can numerically find the energy spectrum of the Hamiltonian (\ref{kiraevham}) for open boundary conditions. \\
To study Majorana zero modes in our non-Hermitian system, we transform the Hamiltonian (\ref{kiraevham}) from the operators $a_j$ to the Majorana operators, which are Hermitian fermionic operators that square to $\ds{1}$.
\begin{figure}[t]\label{20}
\includegraphics[width=8.0cm]{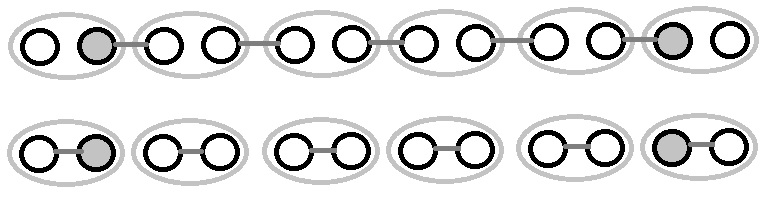}
\caption{ Two phases of the Kitaev chain with $N=12$. The shaded elements of the chain represent gain and loss ($\ds{g_j=g_0(\delta_{j,2}-\delta_{j,11})}$). In the topological phase (top) unpaired Majorana fermions are located on edges. The bottom figure is the same lattice but in the trivial phase. }
\end{figure}
\begin{eqnarray}\label{majodef}
{\gamma}_{j,A} =a_j^{\dagger}+a_j;~~{\gamma}_{j,B} =i(a_j^{\dagger}-a_j)
\end{eqnarray}
The fermion anti-commutation relation for Majorana fermions is given by $\ds{  \{  \gamma_{j,k} ,\gamma_{j^{\prime},k^{\prime}}    \}=2\delta_{j,j^{\prime}}  \delta_{k,k^{\prime}}}$. The Majorana fermion operators do not obey the usual Pauli principle of fermions since $\ds{{\gamma}^{\dagger}_{j,A}={\gamma}_{j,A}}$ and $\ds{{\gamma}^{\dagger}_{j,B}={\gamma}_{j,B}}$. Instead, the Majorana operators have the property,  $\ds{ {\gamma}^2_{j,k}=1}$. Therefore, we see that a Majorana fermion is its own anti-particle. Let us rewrite the Hamiltonian (\ref{kiraevham}) in the Majorana basis using the above transformation in the limit $\ds{\mu=0}$ and $\ds{\Delta=T\neq0}$. In this case, the existence of stable Majorana zero modes can be clearly seen. It is given by
\begin{eqnarray}\label{kiraevhamreduced}
H=-iT\sum_{j=1}^{N-1} {\gamma}_{j,B} \gamma_{j+1,A}-\frac{1}{2}\sum_{j=1}^{N}   g_{j}  {\gamma}_{j,A} \gamma_{j,B}+\frac{i}{2}\sum_{j=1}^{N}  g_j
\end{eqnarray}
It is now clear why we impose two conditions on the non-Hermitian strength. Firstly, the last constant term in (\ref{kiraevhamreduced}) is purely imaginary and shifts the whole spectrum. This term vanishes since we already assumed that gain and loss are balanced in the system. Secondly, assuming that $\ds{g_1=g_N=0}$ makes the transformed Hamiltonian interesting in the sense that the Majorana operators $\ds{\gamma_{1,A} }$ and $\ds{\gamma_{N,B} }$ are explicitly absent from the Hamiltonian (they commute with the Hamiltonian $\ds{  \left[  H,\gamma_{1,A} \right] = \left[  H,\gamma_{N,B} \right] =0}$). As a result, occupying these states requires zero energy. Because of the absence of these two unpaired Majorana operators, we say that Majorana zero mode operators, or simply Majorana zero modes, appear in our non-Hermitian system. We emphasize that this is true even if gain and loss are randomly distributed in the chain as long as gain and loss are balanced. In other words, $\mathcal{PT}$ symmetry is not a necessary condition for the reality of Majorana zero modes. This doesn't mean that all the energy eigenvalues are real valued. In fact, they are either completely or partially real valued. However, what it is clear is that Majorana modes have real valued energy (exactly zero energy). This can be understood as follows. Since no gain and loss are present on edges, Majorana zero modes (localized exactly on edges) experience no gain and loss. This leads to stable Majorana modes. If, on the other hand, gain and loss is not balanced in the system, then the Majorana modes have purely imaginary eigenvalues with the energy $\ds{i/2\sum_{j=1}^N g_j}$ as can be seen from (\ref{kiraevhamreduced}). This is interesting in the sense that Majorana modes that are localized on gainless and lossless edges feel the net amount of gain and loss in the whole system. Note that Majorana zero modes must come in pairs since each of them is, in a sense, half a fermion. In other words, a fermion is equivalent to two spatially separated Majorana fermions. We say that $\ds{{\gamma}_{1,A}}$ and $\ds{{\gamma}_{N,B}}$ are localized exactly at the two edges and thus there exists highly non-local entanglement. To this end, we also check our treatment numerically. We consider random values of non-Hermitian strengths along the lattice such that the total gain and loss is zero in the system. We also assume that there exists no gain and loss on edges. We find that the spectrum is partially complex and there exist zero energy solution that corresponds to localized eigenfunctions on edges. \\
\begin{figure}[t]\label{20}
\includegraphics[width=6.0cm]{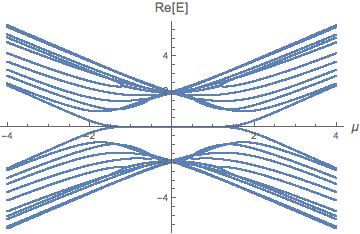}
\includegraphics[width=6.0cm]{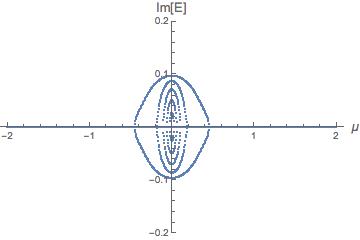}
\includegraphics[width=6.0cm]{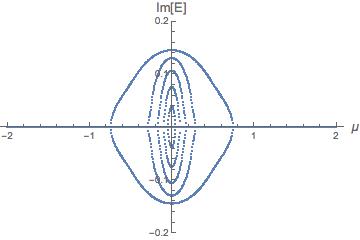}
\includegraphics[width=6.0cm]{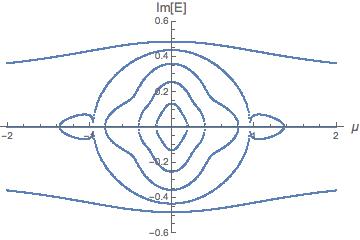}
\caption{ The parameters for all figures are given by $N=12$, $\Delta=T=1$. The top one shows the real part of the spectrum as a function of the chemical potential when $g_0=0.15$. In fact, the real part of the spectrum changes negligibly with $g_0$. As can be seen, zero energy states are available when $-2<\mu<2$. The second, third and bottom figures plot the imaginary part of the spectrum for $g_0=0.1$, $g_0=0.15$ and $g_0=0.5$, respectively. Increasing $g_0$ increases the critical value of chemical potential with which the spectrum becomes real. If $g_0$ is large enough, then the spectrum becomes complex at any $\mu$. }
\end{figure}
\begin{figure}[t]\label{20}
\includegraphics[width=6.0cm]{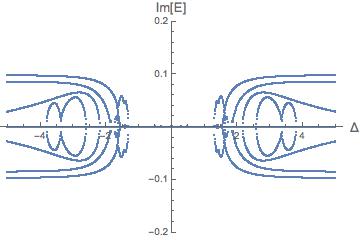}
\includegraphics[width=6.0cm]{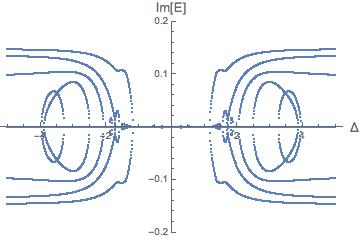}
\includegraphics[width=6.0cm]{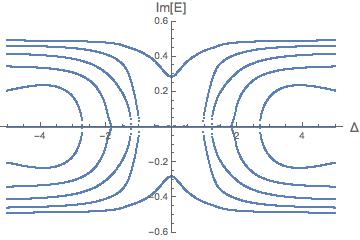}
\caption{ The figures show the imaginary parts of spectra as a function of $\Delta$ for the parameters $N=12$, $\mu=T=1$. The top, middle and bottom figures are for $g_0=0.1$, $g_0=0.15$ and $g_0=0.5$, respectively. The spectrum is real unless $\Delta$ exceeds a critical value if the non-Hermitian strength is weak. The energy eigenvalues are partially complex if the non-Hermitian strength is strong.}
\end{figure}
Suppose next that $\ds{\mu{\neq}0}$ but $\ds{|\mu|{<}2T}$. The Majorana zero-modes are no longer simply given by $\ds{\gamma_{1,A} }$ and $\ds{\gamma_{N,B} }$. Therefore, wave functions that correspond to spatially separated Majorana modes are not exactly localized on edges. In fact they decay exponentially along the chain and have a negligible overlap. Let us now study the reality of spectrum of these Majorana modes by numerically solving our system for a chain with $N=12$ sites and open boundary conditions. We suppose $T=1$ without loss of generality. We consider $\ds{g_1=g_N=0}$ and $\ds{g_j}=(-1)^j g_0$, otherwise, where $\ds{g_0}$ is a constant. In this way, alternating gain and loss occurs through the lattice except on edges. The non-Hermitian part of the Hamiltonian is $\mathcal{PT}$ symmetric if $N$ is an even number. Recall that the full Hamiltonian is not $\mathcal{PT}$ symmetric in the usual sense. We numerically find that the spectrum is real in a broad range of the parameters $\ds{\mu}$ and $\ds{\Delta}$. This interesting result predicts the existence of Majorana modes with real spectrum in our non-Hermitian system. The fig-2 and fig-3 show the imaginary part of the spectrum as a function of $\ds{\mu}$ and $\ds{\Delta}$ for fixed $\ds{\Delta}$ and $\ds{\mu}$, respectively. As we have discussed above, the system is topologically nontrivial when $\ds{-2<\mu<2}$. In the limit $\ds{\mu=0}$, the spectrum is partially complex but the Majorana modes have real energy eigenvalues. As can be seen from the Fig-2, the imaginary part of the spectrum vanishes at a critical $\ds{\mu}$ that depends on both $\ds{g_0}$ and $\ds{\Delta}$. The critical value of $\ds{\mu}$ is smaller than $2$ in most cases. This means that the energy eigenvalues are completely real in the topologically nontrivial region if $\mu$ is beyond a critical number. As a result, we say that stable Majorana mode that exponentially decay along the chain appear in our system. The reality of the spectrum is remained even the system enters topologically trivial region, where $\ds{|\mu|>2}$. Note that the two phases are distinguished by the presence or absence of unpaired Majorana zero modes. The above discussion holds if non-Hermitian strength is weak. If it exceeds a critical value, $\ds{g_0>g_c}$, then the spectrum becomes complex in the whole region as shown the in the last panel of the fig-2.  \\
One can also fix $\ds{\mu}$ and investigate the reality of the spectrum as a function of $\ds{\Delta}$. The fig-3 plots it for three different values of $g_0$. The system admits real spectrum for fixed $\ds{\mu}$ unless $\ds{\Delta}$ exceed a critical value for weak values of $g_0$. If $g$ exceeds a critical value then the energy eigenvalues are all complex valued. \\
To sum up, we have studied a non-Hermitian generalization of the Kitaev model with balanced gain and loss. We have shown that stable Majorana zero energy modes exist in the zero chemical potential limit even if gain and loss are randomly distributed in the chain. If the chemical potential is different from zero, we considered a $\mathcal{PT}$ symmetric distribution of gain and loss and we have found stable Majorana zero modes. In this case, we have shown that the system admits real spectrum as long as chemical potential and non-Hermitian strength don't exceed critical values. In all cases, the Hamiltonian is not $\mathcal{PT}$ symmetric.

\end{document}